\documentclass[aps,prl,twocolumn,superscriptaddress,showpacs]{revtex4}

\usepackage{graphicx}

\renewcommand{\>}{\rangle}
\def\va{\vec{\alpha}}

\def\u0{\underline{0}}

\def\tx{\tilde{x}}
\def\txd{\dot{\tilde{x}}}

\begin{document}

\title{Boosting search by rare events}

\author{Andrea Montanari}
\email[]{montanar@lpt.ens.fr}
\affiliation{Laboratoire de Physique Statistique de l'ENS, 24 rue
Lhomond, 75231 Paris cedex 05, France}

\author{Riccardo Zecchina}
\email[]{zecchina@ictp.trieste.it}
\altaffiliation{Permanent address: ICTP,
Strada Costiera 11, I-34100 Trieste, Italy.}
\affiliation{Laboratoire de Physique Th\'eorique et Mod\`eles
Statistiques, Universit\'e Paris Sud, 91405 Orsay, France}

\begin{abstract}

Randomized search algorithms for hard combinatorial problems exhibit a large
variability of performances. We study the different types of rare events
which occur in such out-of-equilibrium stochastic processes and we show how
they cooperate in determining the final distribution of running times. As a
byproduct of our analysis we show how search algorithms are optimized by
random restarts.

\end{abstract}


\maketitle

Recent years have witnessed an increasing convergence of research themes
coming from out-of-equilibrium statistical physics and computer science or
discrete mathematics \cite{TCS,Codes,SpinGlasses}. For instance, giving
a 'static' characterization of systems displaying an extremely slow dynamics
is a central problem both in computer science~\cite{nature} and in spin glass
theory~\cite{review_dynamics}. The results in these fields are strongly
focused on the {\it typical} properties of large random systems. This approach
is justified as long as the quantities of interest concentrate in probability
around some typical value when the size diverges (the so called self-averaging
property).

In this letter we provide an analytical and numerical study of different types
of rare events which occur in the time evolution of randomized search
algorithms for hard optimization problems. As a byproduct of our analysis, we
find a general picture for understanding and optimizing the introduction of
restarts in randomized search algorithms. This has recently proven to be an
highly effective technique for improving such algorithms \cite{Heavy_tail}.

It is a well known fact (and a basic problem for both theoretical and applied
computer science) that the so called NP-complete~\cite{GaJo} combinatorial
decision problems might require computational resources that grow
exponentially with the number of variables $N$ needed for their encoding.
However combinatorial search methods often exhibit remarkable variability in
performance: it is not uncommon to observe a combinatorial method ``hang'' on a
given instance of a problem, whereas a different heuristic algorithm, or even
just another stochastic run, solves the instance quickly.

With the aim of clarifying such behavior, in the recent years there has been
an intense research activity on randomly generated hard combinatorial problems
which has lead to the identification of non-trivial problem
ensembles~\cite{AI}. Particularly representative and widely studied examples
are satisfiability of random Boolean expressions, vertex coloring and covering
of random graphs and number partitioning~\cite{TCS}.

This type of setting gives us much more freedom than in a standard physical
experiment. Indeed an algorithm can be run an exponential number of times with
each run in turn possibly taking exponential time. In such a situation rare
events may have dramatic effects and completely determine the total
computational time and the outcome of such random-restart experiments. We will
show that there exist distinct sources of hardness fluctuations, static (i.e.
intrinsic) and dynamic (algorithm-dependent), which account for the
variability of resolutions times.

While our approach is general and applies to a wide class of problems, in what
follows we focus on the (NP-complete) vertex cover (VC) problem restricted
over random graphs. The choice of VC is dictated mainly by its relative
simplicity. As expected, extensive numerical experiments match the analytical
predictions.

The action of a backtrack algorithm on combinatorial problems resembles
decimation flows in statistical mechanics \cite{CoMo}. 
The algorithm proceeds by choosing
at random one or more variables at a time and assigning their values according
to some heuristics. The problem is then turned into a sub-problem in which the
assigned variables act as correlated quenched randomness. This evolution can
be described macroscopically by keeping track of a proper set of average
quantities $\vec{\alpha}$ (e.g. the ratio of the number of non-satisfied
constraints to the number of non-assigned variables). The sub-instance
generated by fixing a fraction $t$ of the $N$ variables may however not have
any solution (this happens because the algorithm made a wrong assignment at an
earlier stage). Sooner or later the algorithm detects the inexistence of
solutions compatible with the variables assigned so far and begins a backtrack
correction process which may take an exponential number of steps to correct
early mistakes.

The fraction $t$ of assigned variables acts therefore as a control parameter
and the system undergoes a SAT/UNSAT phase transition (i.e. a transition from
a satisfiable to an un-satisfiable instance of the problem) when $t$ crosses
some critical value $t_c$ \cite{CoMo}. This corresponds to the trajectory
$\vec{\alpha}(t)$ crossing a critical surface in the static phase diagram at
$\vec{\alpha}_c\equiv \vec{\alpha}(t_c)$. For simple enough randomized
algorithms it is possible to compute the size $e^{N\Omega(\vec{\alpha})}$
\cite{CoMo,WeHa2} of the backtracking tree for an UNSAT instance characterized
by the parameters $\vec{\alpha}$. The computational complexity of the
algorithm is therefore given by $\exp[(1-t_c)\Omega(\vec{\alpha}_c) N]$.

How do rare events enter this scenario and how to take advantage
of their presence? There are at least two competing phenomena
involving large deviations which affect the resolution time.

{\bf (I)} Let us assume that the trajectory $\va(t)$ follows the most probable
line. Once the SAT/UNSAT critical line is crossed there still exists a small
probability $\exp[-N(1-t)\psi(\va(t))]$ of having generated a subproblem which
is solvable. The deeper one goes into the UNSAT phase the smaller will be such
probability. On the other hand such a rare event corresponds to a reduction of
order $N$ of the size of the problem and, therefore, to an exponential
reduction of the size $\exp[N(1-t)\Omega(\va(t))]$ of the backtracking tree.

This trade-off can be exploited in a random restart algorithm: we interrupt
the search after $e^{N\tau_R}$ backtracking steps and re-run it (with
different random numbers). The probability of finding a solution in one of
such runs is given by $P_S\approx \exp[-N\min_t (1-t)\psi(\va(t))]$, where $t$
is constrained by the fact that the size of the backtracking tree must be
smaller than $e^{N\tau_R}$: this implies $(1-t)\Omega(\va(t))\le \tau_R$.
Assuming that different stochastic runs quickly lead to uncorrelated
sub-problems, a solution is found after $N_R\approx 1/P_R$ restarts. The
complexity of the algorithm is therefore $\exp[N\tau(\tau_R)]$, where
\begin{eqnarray}
\tau(\tau_R) = \tau_R +\min_t\, (1-t)\psi(\va(t))\, ,
\label{Static}
\end{eqnarray}
and where the minimizing value of $t$ must satisfy 
\begin{eqnarray}
(1-t)\Omega(\va(t))\le \tau_R .
\label{time_contraint}
\end{eqnarray}

{\bf (II)} The above scenario is however largely incomplete. Indeed there
exist another, dynamical, source of fluctuations: $O(1)$ fluctuations with
respect to the typical trajectory (this effect has been studied and 
pointed out to us by S.~Cocco and R.~Monasson \cite{CoMoII}). 
At time $t$ the macroscopic parameters take
the value $\va$ with probability $\exp[-N I_t(\va)]$ (with $I_t(\va)=0$ along
the typical trajectory $\va = \va(t)$). Again, such a rare event implies an
exponential change in the computational complexity, and the possible gain can
be exploited by the random restart algorithm. Equation (\ref{Static}) must be
properly generalized. We get
\begin{eqnarray}
\tau(\tau_R) = \tau_R +\min_{t,\va}\, \left\{ I_t(\va)+(1-t)\psi(\va)\right\}
\, ,
\label{Dynamic}
\end{eqnarray}
always with the constraint Eq.(\ref{time_contraint}).
These are not the only sources of fluctuations but they give 
a quite accurate picture of the phenomenon.

Let's apply the above scheme to the case of VC, which is at the same time
NP-complete \cite{GaJo} and very easy to define: Consider an undirected
graph $G=(V,E)$ with $N$ vertices $i\in V=\{1,2,...,N\}$ and $L$ edges
$(i,j)\in E\subset V\times V$. The problem consists in distributing $X$
covering marks over the vertices in such a way that every edge of the graph is
covered, that is it has at least one of its ending vertices which is marked.
If such covering can be found the graph is said to be coverable (COV),
otherwise it is uncoverable (UNCOV).

A non trivial ensemble of graphs which captures some relevant computational
features of VC at the level of typical or average cases, is the set of random
graphs $G_{N,L}$ with $N$ vertices and $L$ edges (and flat probability
distribution). Similarly to other random NP-complete problems~\cite{TCS}, a
threshold phenomena occurs as the control parameter $X$ is
changed\cite{WeHa1}. For a given average connectivity $c=2L/(N-1)$, when the
number $X=xN$ of covering marks is lowered the model undergoes a COV/UNCOV
transition at some critical density of covers $x_c(c)$ for $N\to \infty$.
Statistical mechanics methods allow for a precise estimate of $x_c(c)$
\cite{nota0} and probabilistic tools provide rigorous lower and upper bounds
for such a threshold \cite{Ga,Fr}. For $x>x_c(c)$, vertex covers of size $Nx$
exist with probability one, for $x<x_c(c)$ the available covering marks are
not sufficient. The statistical mechanics analysis is performed by mapping the
optimization problem onto a zero temperature disordered system with
Hamiltonian$ $
\begin{eqnarray}
H(\{n\}) =  \sum_i \delta_{n_i,0}\, , 
\label{Model}
\end{eqnarray}
where $n_i\in\{0,1\}$ ($n_i=0$ if a mark is put on vertex $i$) and satisfy an
excluded volume constraint: if $(ij)\in G$ then either $n_i=0$ or $n_j=0$. The
ground state energy $E_{GS}$ of the model is the minimum number of marks
needed for covering the graph: for $X\ge E_{GS}$ the graph is COV, while for
$X< E_{GS}$ it is UNCOV.

It is known experimentally and analytically for some algorithms \cite{WeHa2}
that the typical computational cost, given e.g. by the number of visited
decision nodes in the backtracking tree, becomes exponential for initial
conditions in a region close or below $x_c(c)$, while it remains linear well
inside the coverable phase, $x>x_c(c)$. This easy-hard scenario characterizes
the typical-case complexity pattern found in other NP-complete random
ensembles \cite{AI}.

We consider the following backtrack algorithm~\cite{Rag}. During
the computation, a vertices can be {\em covered}, {\em uncovered} or just {\em
free}, meaning that the algorithm has not yet assigned any value to that
vertex. Here is the recursive procedure: The algorithm chooses a vertex $i$ at
random among those which are {\em free} (at $t=0$ all vertices are {\em
free}). If $i$ has neighboring vertices which are either {\em free} or {\em
uncovered}, then the vertex $i$ is declared {\em covered} first. In case $i$
has only covered neighbors, the vertex is declared {\em uncovered}. The
process continues unless the number of covered vertices exceeds $Nx$.
If the algorithm backtracks, then the opposite choice is taken for the vertex
$i$ unless this corresponds to declaring {\em uncovered} a vertex whose
neighbors are all {\em uncovered}. The algorithm halts if it finds a solution
(and declares the graph to be COV) or after exploring all the search tree (in
this case it declares the graph to be UNCOV).

\begin{figure}
\includegraphics[width=0.75\columnwidth,angle=-90]{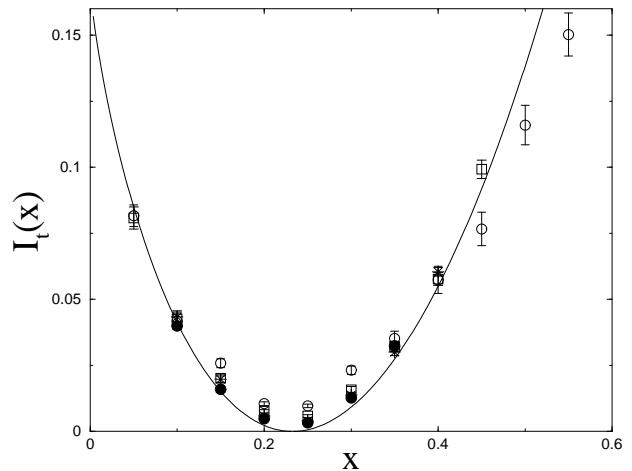}
\caption{Dynamical rare events. We consider the probability 
$P_t(x)$ that, after $Nt$ steps we are left with  $Nx$ marks, and
we plot $I_t(x)= -\log P_t(x)/N$. The continuous line is the theoretical
prediction $I_t(x) = \min_c I_t(c,x)$, while the symbols are
numerical results for $N=100$ (empty circles), $200$ (squares), $300$ 
(stars) and $400$ (full circles). We used the the initial condition $c_0=2$, 
$x_0=0.5$, and $t=0.5$.}
\label{Xfluct}
\end{figure}

In order to study the algorithm we need to follow the variables $X,N,L$ which
become time dependent. In one time step ($T \to T+1$), the
probability for a change $L\to L+\Delta L$ in the number of links and $X\to
X+\Delta X$ in the number of available covering marks reads
\begin{eqnarray}
P_{\Delta L, \Delta X} =  e^{-c}\left[
\delta_{\Delta X,0}\delta_{\Delta L,0} +
\sum_{k=1}^{\infty}\frac{c^k}{k!}\delta_{\Delta X,-1}
\delta_{\Delta L,-k}\right] .
\label{OneStep}
\end{eqnarray}
This defines a Markov process in the space $(X,G_{N,L})$ which mimics the
effects of the algorithm. We want to iterate the above step $\Delta T$ times
and compute the corresponding transition probability. Let us introduce the
rescaled time $t=T/N$ (i.e. the fraction of assigned variables) and the
macroscopic time dependent parameters $c(t) = 2L_T/N_T$ and $x(t)=X_T/N_T$
(which we denoted collectively by $\va$ in the general description of our
approach). Due to the Markovian structure of this process, the probability for
a trajectory $\va(t) \equiv (c(t),x(t))$ can be written in a path integral
form. To the leading order we get $P[c,x] = \int \!\!{\cal D}s\,\exp\{-N
\int\!dt\,{\cal L}_t(c,x,s) \}$, where
\begin{eqnarray}
&{\cal L}_t(c,x,s)=-\frac{i}{2}s \partial_t[(1-t)c]+
(-\txd)\log(-\txd)&\nonumber \\ 
&+(1+\txd)\log(1+\txd)
+\txd\log\left(\exp(ce^{is})-1\right)+c&  ,
\label{Lagrangian}
\end{eqnarray}
where we used the shorthand: $\tx(t)\equiv (1-t) x(t)$. The transition
probability $P_{t_1}(c_0,x_0\to c_1,x_1)$ is given by the corresponding
constrained path-integral. Such an integral can be computed by saddle-point,
leading to an explicit formula for the trajectories:
\begin{eqnarray}
c(t) & = & \frac{c_0}{1-t}-\frac{2}{B(1-t)}\int_{e^{B(1-t)}}^{e^B}\!\!\!\!
d\zeta\; \frac{\log\zeta}{\zeta-A}\, ,\label{SolC}\\
x(t) & = & \frac{x_0-t}{1-t}-\frac{A-1}{AB(1-t)}\log\left\{
\frac{1-Ae^{-B}}{1-Ae^{-B(1-t)}}
\right\} \label{Sol} ,
\end{eqnarray}
where the two integration constants ($A$ and $B$) must be computed
from the conditions $c(t_1)=c_1$, $x(t_1) = x_1$. The large deviation
functional is $I_{t_1}(c_1,x_1) = \int_0^{t_1} \!dt {\cal L}_t(\cdot)$, where
the integral is computed along the trajectory (\ref{SolC}),(\ref{Sol}). For
$A=0$ and $B=c_0$ we recover the typical trajectory \cite{nota1}
and we get $I_t(c,x)=0$.
As shown in Fig. \ref{Xfluct}, numerical simulation are in remarkable agreement
with the analytical predictions.

A subgraph generated according to the process described above can be still COV
(with an exponentially small probability) after the trajectory
$\va(t)=(c(t),x(t))$ has entered the UNSAT phase (i.e. after
$x(t)<x_c(c(t))$). Repeated restarts can exploit this rare event. The size
$\exp[N(1-t)\Omega(\va(t))]$ of the backtrack tree at any point in the UNCOV 
region can be computed analytically \cite{WeHa2} and used in
Eq.(\ref{time_contraint}). Hence, in order to evaluate Eq.(\ref{Dynamic}), we
just need to compute the probability of being COV in the UNCOV region, that is
we need to know the probability distribution of the ground state energy of the
model (\ref{Model}). This computation can be carried over by the replica
method. We notice that the replicated partition function averaged over the
disorder reads
\begin{eqnarray}
\<Z^n\>\to\int_0^{\infty}\!\!dP(E_{GS})\, e^{-\omega E_{GS}}\, ,
\end{eqnarray}
where one takes the zero temperature limit $\beta \to \infty$ keeping $\omega
\equiv n\beta$ fixed. As $N\to\infty$, $\<Z^n\>\sim \exp [-N\phi(\omega)]$ and
$P(E_{GS}) \sim \exp[-N\psi(x)]$, where $x = E_{GS}/N$. We get therefore
$\phi(\omega) = \left.\psi(x)+\omega x
\right|_{\omega = -\partial_{x}\psi}$.

\begin{figure}
\includegraphics[width=0.75\columnwidth,angle=-90]{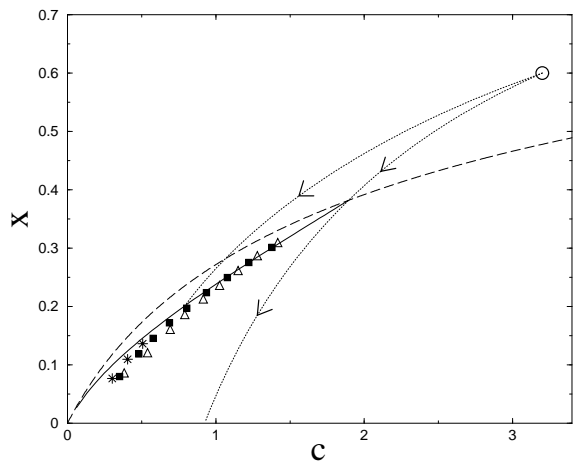}
\caption{Random restart experiments with initial condition at $c=3.2$, $x=0.6$
(empty circle). The long dashed line is the replica symmetric critical line
$x=x_c(c)$. The rightmost dotted line is the typical trajectory. The leftmost
one is the rare trajectory followed by the last (successful) restart of the
algorithm when $\tau_R=0.1$. The symbols are numerical results for the
$(c,x)$-values corresponding to the backtrack tree generated by the
algorithm since the last restart. Triangles, squares and stars correspond,
respectively, to $N=30$, $60$, $120$. The continuous line is the theoretical
prediction for the same quantity (i.e. the minimizing $\va\equiv (c,x) $ in Eq.
(\ref{Dynamic})).}
\label{cxfig}
\end{figure}

The small $\omega$ behavior of $\phi(\omega)$ yields the typical ground state
energy and its small fluctuations. The knowledge of the whole function
$\phi(\omega)$ gives the large deviation properties of the ground state
energy. In general $\psi(x)$ is convex and has its unique zero at the typical
ground state energy $x =x_c(c)$. The probability that a graph in the ensemble
is coverable with $X=Nx<Nx_c(c)$ marks is given by $\exp[-N\psi(x)]$.
A replica symmetric calculation yields
\begin{eqnarray}
\phi(\omega) & =& c(1-F_{Q}) +\frac{c}{2}\log F_{Q}-\\
&& -\log[e^{-\omega}+(1-e^{-\omega})e^{-cF_{Q}Q}]\, ,\nonumber
\end{eqnarray}
where we used the short-hand $F_Q = [1+(e^{-\omega}-1)Q^2]^{-1}$ and $Q$
satisfies the self-consistency equation: $Q^{-1}=1-e^{-\omega}[1+\exp(cF_QQ)]$
. Figure \ref{cxfig} gives the geometrical picture of a random restart
experiment. Quite remarkable is the prediction on the $(c,x)$-values up to
which the algorithm has to backtrack before finding the solution. Such a curve
lies well inside the UNCOV region indicating that the two types of rare events
are both relevant for $\tau_R>0$.
\begin{figure}
\includegraphics[width=0.75\columnwidth,angle=-90]{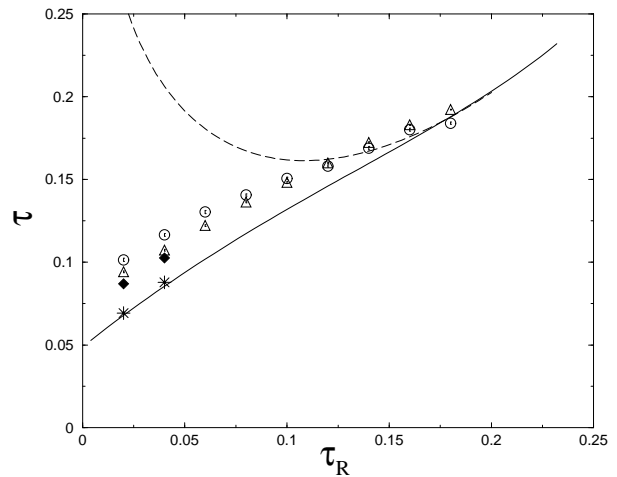}
\caption{Typical computational complexity of the random restart algorithm.
Here we plot the logarithm of number of nodes visited by the algorithm
divided by the size $N$, for different values of the restart rate 
$\tau_R$. Symbols refer to $N=30$ (circles), $60$ (triangles),
and $120$ (diamonds).   The stars are the result
of an $N\to\infty$ extrapolation.
The continuous and dashed lines reproduce the theoretical 
prediction with, cf. Eq. (\ref{Dynamic}), and without,
cf. Eq. (\ref{Static}), dynamical rare events.\vspace{-0.1cm}}
\label{cplx}
\end{figure}
In Fig. \ref{cplx} we consider the computational complexity
$e^{N\tau(\tau_R)}$ of the random restart algorithm for the initial condition
$c=3.2$, $x=0.6$. Finite size effects are important for the achievable sizes of
the problem. An extrapolation can be done for the smaller values of $\tau_R$,
where we were able to run the algorithm on much larger systems.

Building on large deviations results we have shown that running times of
randomized complete search algorithms can be greatly reduced by a restart
strategy. The optimal restart rate $\tau^{\rm opt}_R$ can be computed within
our approach: for VC we find $\tau^{\rm opt}_R=0$.
This result highlights the relevance of the sub-exponential  regime
which has been investigated thoroughly in Ref. \cite{CoMoII}.
In more general cases we expect $\tau^{\rm opt}_R>0$ \cite{Heavy_tail}. 
 In would be interesting to explore whether this rare events
scenario also applies to other classes of stochastic processes, both
algorithmic and physical.

Credit for this paper should be shared with S.~Cocco and R.~Monasson
for very fruitful conversations concerning dynamical fluctuations 
in random 3-SAT\cite{CoMoII}. We also
greatly thank A.~Braunstein, S.~Franz, M.~M\'ezard, G.~Parisi, N.~Sourlas
and M.~Weigt.


\begin{thebibliography}{99}
\bibitem{TCS} O. Dubois, R. Monasson, B. Selman and R. Zecchina (eds.), 
Theor. Comp. Sci. \textbf{265}, issue 1-2 (2001).
\bibitem{Codes} H. Nishimori, {\it Statistical Physics of Spin Glasses and
Information Processing}, Oxford University Press, 2001
\bibitem{SpinGlasses} M. M\'ezard, ``Theory of random solid states'',
unpublished.
\bibitem{nature} R. Monasson, R. Zecchina, S. Kirkpatrick, B. Selman, and L.
Troyansky, Nature \textbf{400}, 133 (1999).
\bibitem{review_dynamics} 
J.-P. Bouchaud, L.F. Cugliandolo, J. Kurchan and M. M\'ezard, in 
{\it Spin Glasses and Random Fields},A.P. Young ed., World
Scientific (1997).
\bibitem{Heavy_tail} C.P. Gomes, B. Selman, N. Crato, K. Kautz, J. of
Automated Reasoning \textbf{24}, 67 (2000); C.P. Gomes, B. Selman, and H.
Kautz. Proc. AAAI-98, 431--437, Madison, WI, July 1998;
\bibitem{GaJo} M. R. Garey and D. S. Johnson, {\it Computers and
intractability} (Freeman, New York, 1979).
\bibitem{AI} T. Hogg, B. A. Huberman, and C. Williams (eds.), 
{\it Artif. Intell.} {\bf 81} (I+II) (1996)
\bibitem{Rag} R. Motwani, P.Raghavan, {\it Randomized Algorithms}, 
Cambridge University Press, (2000)
\bibitem{CoMo} S. Cocco and R. Monasson,{\it Phys. Rev. Lett.} {\bf 84}, 1654
(2001); {\it Eur. Phys. J.} {\bf B 22}, 505 (2001). 
\bibitem{WeHa2} M. Weigt and A. K. Hartmann, {\it Phys. Rev. Lett.} {\bf 86},
1658 (2001)
\bibitem{nota0} For $c<e$, the statistical mechanics analysis\cite{WeHa1}
gives $x_c(c)= 1- (2W+W^2)/2c$, where $We^W=c$. For
$c>e$ there appear replica symmetry breaking effects.
\bibitem{Ga} P. G. Gazmuri, {\it Networks} {\bf 14}, 367 (1984); 
\bibitem{Fr} A. M. Frieze, {\it Discr. Math.} {\bf 81}, 171 (1990)
\bibitem{WeHa1} M. Weigt and A. K. Hartmann, {\it Phys. Rev. Lett.} 84, 6118
(2000)
\bibitem{nota1} $c(t)=c_0(1-t)$, $x(t)= \{
x_0-t+\frac{1}{c_0}[e^{-c_0(1-t)}-e^{-c_0}]\}/(1-t)$, see \cite{WeHa1}
\bibitem{CoMoII}  S. Cocco and R. Monasson, submitted

\end{thebibliography}
\end{document}